\documentclass[12pt,amsfonts]{article}

\usepackage{amsfonts}

\def\vp{\varphi}

\def\half{\textstyle{\frac{1}{2}}}

\def\H{{\cal H}}

\def\p{\varphi}

\def\H{{\cal H}}

\def\l{\lambda}

\def\t{\textstyle}

\def\ra{\rightarrow}
\def\tint{{\textstyle\int}}

\def\d{\partial}

\def\b{\begin{eqnarray*}}  
\def\e{\end{eqnarray*}}    
\def\bn{\begin{eqnarray}}  
\def\en{\end{eqnarray}}   

\def\<{\langle}
\def\>{\rangle}

\def\bk{\mathbf k}
\def\bm{\mathbf m}

\def\de{\delta}
\def\no{\nonumber}

\def\ds{d^s\!x}
\def\k{\kappa}

\def\L{\Lambda}

\def\hk{\hat{\kappa}}
\def\{{\lbrace}
\def\hv{\hat{\varphi}}
\def\}{\rbrace}
\begin{document}

\title{An Ultralocal Classical and \\ Quantum Gravity Theory}        
\author{John R. Klauder\footnote{klauder@phys.ufl.edu} \\
Department of Physics and Department of Mathematics \\
University of Florida,   
Gainesville, FL 32611-8440}
\date{ }
\bibliographystyle{unsrt}

\maketitle 

\begin{abstract}

An ultralocal form of any classical field theory eliminates all spatial derivatives in its action functional, e.g., in its Hamiltonian functional density. It has been applied to covariant scalar field theories and even to Einstein's general relativity, by Pilati, as an initial term in a perturbation series that aimed to restore all omitted derivatives. Previously, the author has quantized ultralocal scalar fields by affine quantization to show that these non-renormalizanle theories can be correctly quantized by affine quantization; the story of such scalar models is discussed in this paper. The present paper will also show that ultralocal gravity can be successfully quantized by affine quantization.

 The purpose of this study is that a successful affine quantization of any ultralocal field problem implies that, with properly restored derivatives, the classical theory can, in principle,  be guaranteed a successful result using either a canonical quantization or an affine quantization.
In particular, Einstein's gravity requires an affine quantization, and it will be successful.
\end{abstract}

\section{Ultralocal Scalar Field Theories}
Before considering gravity, it can be useful to review a modest summary of the results of canonical  quantization when it has been used to study a variety of covariant scalar field models.

A traditional covariant scalar field has a classical ($c$) Hamiltonian given by
  \bn H_c=\tint \{\half[\pi(x)^2+({\overrightarrow{\nabla}}\vp)(x)^2+m^2_0\vp(x)^2]+g_0\,\vp(x)^p\}\;\ds \;, \en
  with $p$ an even positive integer, $s$, the number of spacial dimensions, is a positive integer, $m_0^2>0$, and $g_0\geq0$. With $n=s+1$ spacetime dimensions and using canonical quantization, a satisfactory quantization appears for $p<2n/(n-2)$. If $p\geq 4=n$ a Monte Carlo study \cite{3}
  and analysis \cite{1,2} only found `free theory results', while if perturbation series were used when $p>2n/(n-2)$, then only, divergent, non-renormalizable behavior appears.
  The ultralocal ($u$) scalar field model here has a classical Hamiltonian given by
    \bn H_u=\tint \{\half[\pi(x)^2+m_0^2\vp(x)^2]+g_0\vp(x)^p\} \;\ds \;, \en
    and with canonical quantization, a divergent perturbation series for $p>2$ and any $n\geq2$ leads to undesired results. Let us choose a different path.
    
    The domain of $H_u$ consists of all, continuous momentum functions $\pi(x)$ and scalar fields $\vp(x)$ for which
    $0<H_u<\infty$. Our interest is focussed on $p>2$, and thus $p\in\{4,6,8,...\}$, and 
   clearly, for all such $p>2$ and all $s\in\{1,2,3,...\}$, the domain ${\cal{D}}_{p>2}(\pi,\vp)\subset {\cal{D}}_{p=2}(\pi,\vp)$; we will raise the issue of domains again at the end of this story.
    When quantized by using canonical quantization, along with a perturbation series in powers of $g_0$, one encounters multiple infinities. 
    
    Since all derivatives have been removed, what happens at any point of $x$ does not affect  what happens at any other point $x'\neq x$. Just like sums of independent operators in Hamiltonians leads to products of separate and independent wave functions, this implies, for our problem, that the ground state is composed of products at different points, which -- thanks to the Central Limit Theorem -- leads to the characteristic function of the ground state distribution, i.e., the Fourier transformation of the absolute square of the normalized ground state, being given in the form of
       \bn &&C(f)=\int e^{i\tint f(x)\vp(x)\,\ds}\;\Pi_x\,|\Psi_0[\vp(x)]|^2\;d\vp(x) \no \\
       &&\hskip2.5em=e^{-\tint W(f(x))\,\ds}\;,\en
       for a suitable function $W(f(x))$, with $W(f(x))=0$ whenever $f(x)=0$. 
       
       \subsection{An affine ultralocal scalar field}
       Affine classical variables are given by $\k(x)\equiv \pi(x)\,\vp(x)$  and $\vp(x)$, with
       the restriction that $\vp(x)\neq0$, and the Poisson bracket is given by 
       $\{\vp(x),\k(x')\}=\delta^s(x-x')\vp(x)$. The classical ultralocal Hamiltonian expressed in affine 
       variables is given by 
         \bn H_u=\tint \{\half[\k(x)^2\,\vp(x)^{-2}+m_0^2\,\p(x)^2]+g_0\,\vp(x)^p\}\;\ds\;.\en
         
         The basic quantum operators are $\hv(x)\neq0$ and $\hk(x)$, and their commutator is given by
         $[\hv(x),\hk(x')]=i\hbar\delta^s(x-x')\hv(x)$. The quantum affine Hamiltonian is given by
           \bn \H_u=\tint\{\half[\hk(x)\hv(x)^{-2}\hk(x)+m_0^2\,\hv(x)^2]+g_0\,\hv(x)^p\}\;\ds\;, \en
        and the Schr\"odinger representation is given by $\hv(x)=\vp(x)$ and 
        \bn \hk(x)= -\half i\hbar[\vp(x)(\de/\de\vp(x)))+(\de/\de\vp(x))\vp(x)]\:.\en
        
        Clearly, this is a formal equation for the Hamiltonian operator, etc. Such expressions deserve 
        a regularization of these equations.

       \subsection{A regularized affine ultralocal scalar field}
       Our regularization is of the underlying space in which $x\ra \bk\,a$, where $\bk\in\{..., -1,0,1,2,3,...\}^s$ and $a>0$ denotes the tiny distance between lattice rungs. The regularized 
       classical ultralocal Hamiltonian is given by
         \bn H_u={\t\sum}_{\bk} \{\half[\pi_\bk^2+ m_0^2\,\vp_\bk^2]+g_0\,\vp_\bk^p\}\;a^s \;.\en
       The classical affine regularization involves $\k_\bk=\pi_\bk\,\vp_\bk$ and $\vp_\bk$, with
       $\vp_\bk\neq0$, with a Poisson bracket $\{\vp_\bk, \k_\bm\}=\delta_{\bk,\bm}\,\vp_\bk$, and 
       the classical affine regularized ultralocal Hamiltonian is given by
   \bn \H_u= {\t\sum}_{\bk} \{\half[\k_\bk^2\vp_\bk^{-2}+m_0^2\,\vp_\bk^2]+g_0\,\vp_\bk^p\}\;a^s \;.\en
   
         The regularized basic quantum Schr\"odinger operators are given by $\hv_\bk=\vp_\bk$ and
           \bn &&\hk_\bk=-i\half\hbar[ \vp_\bk(\d/\d\vp_\bk)+(\d/\d\vp_\bk)\vp_\bk] a^{-s} \no \\
           &&\hskip1.36em=-i\hbar[\vp_\bk(\d/\d\vp_\bk)+1/2]a^{-s} \;. \en
           An important result is that $\hk_\bk\,\vp_\bk^{-1/2}=0$. The Schr\"odinger equation becomes
  \bn i\hbar \,\d\psi(\vp,t)/\d t={\t\sum}_\bk\{\half[\hk_\bk\vp_\bk^{-2}\hk_\bk+m_0^2\,\vp_\bk^2]+g_0\,\vp_\bk^p \}\;a^s\;\psi(\vp,t)\;. \en
  The normalized ground state of such an equation is given by
      \bn \psi_0(\vp)= \Pi_\bk e^{-V(\vp_\bk)/2}\,(ba^s)^{1/2}\,\vp_\bk^{-(1-2ba^s)/2}\;, \en
      for some real function $V(\vp_\bk)$. Finally, we ask what is the characteristic function for such an equation, and the answer is given by
       \bn &&C(f)=\lim_{a\ra0}\Pi_\bk\,\tint \;e^{i f_\bk \vp_\bk}\; e^{-V(\vp_\bk)}\,(ba^s)\,|\vp_\bk|^{-(1-2ba^s)}\;d\vp_\bk \no \\
       &&\hskip2.53em  =\lim_{a\ra0}\Pi_\bk \{1-(ba^s)\tint[1-e^{if_\bk\vp_\bk }]\,e^{-V(\vp_\bk)}\,|\vp_\bk|^{
       -(1-2ba^s)}\;d\vp_\bk \} \no \\
       &&\hskip2.55em  =\exp\{-b\tint \ds\tint[1-e^{i\,f(x)\l}] e^{-v(\l)}\,d\l/|\l|\} \;.\label{x}\en
       Here $\vp_\bk\ra\l$, and $V\ra v$ to account for changes that may have arisen in $V$ as $a\ra0$.
       The resultant expression in (\ref{x}) is a (generalized) Poisson distribution, which, besides 
       a Gaussian distribution, is the only other form allowed by the Central Limit Theorem.

       As a crude estimate of the large behavior of $\vp$ for different $p$ values we can  examine
       $(-d^2/dx^2+x^p)\,e^{-|x|^\gamma/\gamma}\simeq (-x^{2\gamma-2} + x^p+ \cdots) e^{-|x|^\gamma/
       \gamma}$. To cancel the largest terms we roughly need to have $\gamma=1+p/2$. 
       If $g_0\ra0$, then $p\ra 2=\gamma$, and thus $v(\l)\ra c\l^2$, with $c>0$. This last fact 
       means that sending $g_0\ra0$ does not lead us 
       to a simple, free Gaussian because the given result reflects the continuity of a smaller domain
       of the interacting model as compared to the strictly larger domain of the 
       truly free theory. 
       
       \subsubsection{The main lesson from ultralocal scalar fields}
       The previous subsection found that an ultralocal scalar field model led to acceptable results 
       when $p\geq2$ and $n\geq2$. For certain covariant scalar field models, we have already 
       observed that acceptable results arise by canonical quantization when  $p<2n/(n-2)$. 
       In view of acceptable results for ultralocal scalar fields when $p\geq2$ and $n\geq2$,
       we predict that an affine quantization for covariant scalar fields leads to acceptable results 
       when $p\geq 2n/(n-2)$. Monte Carlo studies, such as those carried out in \cite{3}, 
       could confirm whether this prediction is true or not.
       
       \section{Ultralocal General Relativity}
        An effort to quantize Einstein's theory of gravity has been examined in several
        articles published by the author; see \cite{4,5,6,8}, with \cite{7}, perhaps, being the 
        strongest effort of them all. In light of those articles, we will present a modest selection 
        of the necessary features for an affine quantization of Einstein's gravity.
        
        The phase space variables in the ADM version of classical general relativity \cite{adm}
        are the metric field $g_{ab}(x) $ and the momentum field $\pi^{cd}(x)$, where
        $a,b,c,d,...=1,2,3$, and which a canonical 
        quantization promotes to basic quantum operators. The positivity requirement that
        $g_{ab}(x)\,dx^a\,dx^b>0$ implies that the momentum operator can not be self adjoint.
        An affine quantization chooses the classical metric $g_{ab}(x)$, which has a positive 
        requirement as before, while the momentum 
        field is replaced by the momentric field $\pi^a_b(x)$ $[\equiv \pi^{ac}(x)\, g_{bc}(x)]$.
        These basic affine variables are promoted to quantum operators, both of which can be
        self adjoint, while the metric operator is also positive as desired.
        
        The ADM classical Hamiltonian, with $g(x)\equiv \det[g_{ab}(x)]>0$, is given by
          \bn H_c=\tint\{ g(x)^{-1/2}[\pi^a_b(x)\pi^b_a(x)-\half\,\pi^a_a(x)\pi^b_b(x)]+
          g(x)^{1/2}\,R(x)\,\}\;d^3\!x\; \en
       where $R(x)$ is the 3-dimensional scalar curvature. 
       
       The term $R(x)$ contains all of the spatial derivative terms and the ultralocal version of the 
       classical Hamiltonian is chosen as
        \bn H_u=\tint\{ g(x)^{-1/2}[\pi^a_b(x)\pi^b_a(x)-\half\,\pi^a_a(x)\pi^b_b(x)]+
          g(x)^{1/2}\,\Lambda(x)\}\;d^3\!x\;. \en
          This expression now has a position-dependent `cosmological constant' in place of the scalar  
          curvature. The term $\Lambda(x)$ (imitating $R(x)$) should be a continuous function that 
          obeys 
          $-\infty<\Lambda(x)<\infty$. When
          this Hamitonian is quantizaed the only variables that are promoted
          to quantum operators are the metric field, $g_{ab}(x)$, and the momentric 
          field, $\pi^c_d(x)$; the classical function
          $\Lambda(x)$ is fixed and not made into an operator.\footnote{Pilati discussed a similar
          model \cite{PP,PP2} in an effort to use such a model as the first term in a perturbation 
          series to restore the proper gravity using canonical quantization.}

          \subsection{An affine quantization of ultralocal gravity}
          The quantum operators are $\hat{g}_{ab}(x)$ and $\hat{\pi}^c_d(x)$, and their Schr\"odinger
           representations are given by $\hat{g}_{ab}(x)=g_{ab}(x)$ and
           \bn \hat{\pi}^a_b(x)=-i\half\hbar[g_{bc}(x)(\de/\de g_{ac}(x))+(\de/\de g_{ac}(x))
           g_{bc}(x)] \;.\en
           The Schr\"odinger equation for the ultralocal Hamiltonian is given by
           \bn &&i\hbar\,\d\,\psi(\{g\},t)/\d t=\tint \{\,\hat{\pi}^a_b(x)\,g(x)^{-1/2}\,\hat{\pi}^b_a
           (x)-\half\hat{\pi}^a_a(x)\,g(x)^{-1/2}\,\hat{\pi}^b_b(x) \no \\
           &&\hskip13em +g(x)^{1/2}\,\Lambda(x) \} \;d^3\!x\;\psi(\{g\},t) \;, \label{g} \en
           where the symbol $\{g\}$ denotes the full metric matrix. In addition, we find that
            $\hat{\pi}^a_b(x)\,g(x)^{-1/2}=0$, which is proved in \cite{7}, just above Eq.~(40).
            Solutions of (\ref{g}) are again governed by the Central Limit Theorem.
         
           As was the scalar case in the previous section, the formal expression for (\ref{g}) 
           needs to be regularized.         
          
          \subsection{A regularized affine ultralocal quantum gravity}
          Much like the regularization of the scalar fields, we introduce a discrete version of the 
          underlying space such as $x\ra \bk a$, where $\bk\in\{...,  -1,0,1,2,3,...\}^3$ and
          $a>0$ is the spacing between rungs in
          which, for the Schr\"odinger representation, $g_{ab}(x)\ra g_{ab\,\bk}$ and 
          $\hat{\pi}^c_d(x)\ra \hat{\pi}^c_{d\,\bk}$ that becomes
             \bn &&\hat{\pi}^c_{d\,\bk}=-i\half\hbar[ g_{de\,\bk} (\d/\d g_{ce\,\bk})+
              (\d/\d g_{ce\,\bk})g_{de\,\bk}]\;a^{-s}\;\no \\
              &&\hskip1.9em
               =-i\hbar[g_{de\,\bk}(\d/\d g_{ce\,\bk})+\delta^c_d/2]\;a^{-s}\:.\en
               Take note that $\hat{\pi}^a_{b\,\bk}\,g_\bk^{-1/2}=0$, 
               where $g_\bk=\det(g_{ab\,\bk})$.
               
               The regularized Schr\"odinger equation is then given by
               \bn i\hbar\,\d \psi(\{g\},t)/\d t={\t\sum}_\bk\{\hat{\pi}^a_{b\,\bk}g_\bk^{-1/2}
               \hat{\pi}^b_{a\,\bk}-\half \hat{\pi}^a_{a\,\bk}g_{\bk}^{-1/2}\hat{\pi}^b_{b\,\bk}\no\\
               +g_\bk^{1/2}\Lambda_\bk\,\}\;a^s\;\psi(\{g\},t)\;,\en
               A normalized, stationary solution to this equation is given by
               \bn \psi_Y(\{g\})=\Pi_\bk\,Y(g_\bk,\L_\bk)\,(ba^3)^{1/2}\,
               g_\bk^{-(1-ba^3)/2}\;.\en
               The characteristic function for such an expression is given by
             \bn &&C_Y(f)=\lim_{a\ra0}\Pi_\bk\tint e^{if_\bk g_\bk}\,|Y(g_\bk,\L_\bk)|^2 (ba^3)
             g_\bk^{-(1-ba^3)}\:dg_\bk\no\\
              &&\hskip3em=\lim_{a\ra0}\Pi_\bk\{1-(ba^3)\tint[1-e^{if_\bk g_\bk}]|Y(g_\bk,\L_\bk)|^2
              g_\bk^{-(1-ba^3)}\;dg_\bk \} \no \\
      &&\hskip3em=\exp\{-b\tint d^3\!x\,\tint[1-e^{if(x)\mu}]\,|y(\mu, \L(x))|^2\,d\mu/\mu \}\;, \en
             where the scalar $g_\bk\ra\mu>0$ and $Y\ra y$ to accommodate
             any change in $Y$ due to $a\ra0$. The final result is a (generalized) Poisson 
             distribution, which obeys the Central Limit Theorem.
             
             The formulation of characteristic functions for 
             gravity establishes the suitability of affine quantization as claimed. Although this 
             analysis was only for an ultralocal model, it nevertheless points to the existence of 
             proper solutions for Einstein's general relativity..

               \subsubsection{The main lesson from ultralocal gravity}
               Just like the success of quantizing ultralocal scalar models, we have also showed that 
               ultralocal gravity  can be quantized using affine quantization. The purpose of solving 
               ultralocal scalar models was to
               ensure that non-renormalizable covariant scalar fields can also be solved using affine 
               quantization. Likewise, the purpose of quantizing an ultralocal version of Einstein's
               gravity shows that we should, in principle, and using affine quantization, be able 
               to quantize the genuine version of Einstein's gravity using affine quantization.
               
               The analysis of certain gravity models with significant symmetry may provide examples 
               that can be completely solved using the tools of affine quantization. 
              For readers 
               interested in the why and the how of affine quantization, perhaps \cite{5} could 
               be a good place to start.

\end{document}